\documentclass[10pt]{iopart}
\expandafter\let\csname equation*\endcsname\relax
\expandafter\let\csname endequation*\endcsname\relax

\usepackage{amssymb}
\usepackage{amsmath}
\usepackage{natbib}
\usepackage{xcolor}
\usepackage{subfigure}
\usepackage{ctable}
\usepackage{graphicx}
\usepackage{makecell}
\usepackage{caption} 
\usepackage[shortlabels]{enumitem}
\usepackage{hyperref}
\usepackage{xspace}
\usepackage{lineno}

\usepackage{makecell}
\usepackage{multirow}
\usepackage{ctable}
\usepackage{ulem} 
\usepackage{bm}

\usepackage{xcolor}
\newcommand{\code}[1]{\texttt{#1}\xspace}
\newcommand{\cref}{}
\begin{document}

\title[SBI and Galaxy Spectra]
{DIGS: Deep Inference of Galaxy Spectra with Neural Posterior Estimation}
\author{Gourav Khullar${}^{1,2,3,4,5}$, Brian Nord${}^{1,2,3}$, Aleksandra \'Ciprijanovi\'c${}^1$, Jason Poh${}^{2,3}$, Fei Xu${}^{2,3}$}

\address{${}^1$Fermi National Accelerator Laboratory, Batavia, IL 60510, USA}
%\address{${}^2$Space Telescope Science Institute, Baltimore, MD 21218, USA}
\address{${}^2$Department of Astronomy and Astrophysics, University of Chicago, Chicago, IL 60637, USA}
\address{${}^3$Kavli Institute for Cosmological Physics, University of Chicago, Chicago, IL 60637, USA}
\address{${}^4$Kavli Institute for Astrophysics \& Space Research, Massachusetts Institute of Technology, 77
Massachusetts Ave., Cambridge, MA 02139, USA}
%\address{${}^5$Mathematics and Computer Science Division, Argonne National Laboratory, Lemont, IL 60439, USA} 
\address{${}^5$Department of Physics and Astronomy and PITT PACC, University of Pittsburgh, Pittsburgh, PA 15260, USA}

\ead{gkhullar@uchicago.edu}

\begin{abstract}

With the advent of billion-galaxy surveys with complex data, the need of the hour is to efficiently model galaxy spectral energy distributions (SEDs) with robust uncertainty quantification. 
The combination of Simulation-Based inference (SBI) and amortized Neural Posterior Estimation (NPE) has been successfully used to analyse simulated and real galaxy photometry both precisely and efficiently.
In this work, we utilise this combination and build on existing literature to analyse simulated noisy galaxy spectra. 
Here, we demonstrate a proof-of-concept study of spectra that is a) an efficient analysis of galaxy SEDs and inference of galaxy parameters with physically interpretable uncertainties; and b) amortized calculations of posterior distributions of said galaxy parameters at the modest cost of a few galaxy fits with MCMC methods. 
We utilise the SED generator and inference framework \code{Prospector} to generate simulated spectra, and train a dataset of 2$\times$10$^6$ spectra (corresponding to a 5-parameter SED model) with NPE. 
We show that SBI -- with its combination of fast and amortized posterior estimations -- is capable of inferring accurate galaxy stellar masses and metallicities. 
Our uncertainty constraints are comparable to or moderately weaker than traditional inverse-modeling with Bayesian MCMC methods (e.g., 0.17 and 0.26 dex in stellar mass and metallicity for a given galaxy, respectively). We also find that our inference framework conducts rapid SED inference (0.9-1.2$\times$10$^5$ galaxy spectra via SBI/SNPE at the cost of 1 MCMC-based fit). 
With this work, we set the stage for further work that focuses of SED fitting of galaxy spectra with SBI, in the era of JWST galaxy survey programs and the wide-field Roman Space Telescope spectroscopic surveys.

\end{abstract}


\submitto{MLST}

\ioptwocol

%% main text
%%%%%%%%%%%%%%%%%%%%%%%%%%%%%%%%%%%%
%%%%%%%%%%%%%%%%%%%%%%%%%%%%%%%%%%%%
\section{Introduction}
\label{intro}

Understanding the mass assembly of galaxies across cosmic time is a major goal of modern extragalactic astrophysics; solving this question sheds light onto a galaxy's underlying formation and evolution mechanism. 
Galaxies are well-characterized by features like stellar mass, chemical composition, dust attenuation, current star formation rate, and the star formation history.
These parameters can be accurately inferred from a galaxy's spectral energy distribution (SED).

Within the last two decades, photometry-based SED fitting has become a pivotal method to measure the above properties. 
Ground-based telescopes have been used extensively for large multi-wavelength galaxy surveys -- e.g., Sloan Digital Sky Survey (SDSS, \citealt{Ahumada_2020}), Dark Energy Survey (DES, \citealt{des2018}), and DESI Legacy Imaging Surveys \citep{Dey_2019} -- producing large high-quality complex datasets. 
However, SED studies relying on photometry alone are subject to many challenges, such as the age-metallicity-dust degeneracy \citep{worthey1994,charlot1999}.
SED fitting using spectra mitigate this challenge significantly, especially to constrain galaxy star formation rates, formation timescales, and metallicities, with measurement of absorption line indices and emission line strengths (e.g., \citealt{worthey1994, leja2019b} and references therein.)

There are several cutting-edge SED-fitting pipelines with Bayesian frameworks that use Markov Chain Monte Carlo (MCMC) methods to infer galaxy properties -- e.g., CIGALE, MAGPHYS, and Prospector \citep{Leja_2017,carnall2019, Leja_2019, johnson21}. 
However, the computational time needed by the fitting algorithms in these frameworks -- e.g., MCMC or nested sampling has been recently is a major bottleneck.
With the next generation of telescopes, like the Vera Rubin Observatory/ Legacy Survey of Space and Time (VRO–LSST; \citealt{lsst_2019}) and the Dark Energy Spectroscopic Instrument (DESI; \citealt{desi2016}), tens of millions of  optical and infrared galaxy photometry and spectra will be measured. 
Highly-resolved spectra have higher information content and require more complex and flexible models for fitting. 
Finally, datasets with spaxels (or spatial and spectral pixels) are increasing in number, e.g., data units in integral-field-unit (IFU) spectrograph observations, with JWST IFU spectroscopy of a gravitationally lensed galaxy \citep{khullar_jwst_cj1241_prop}, or the SDSS-MaNGA survey observations of star forming galaxies.

%%%%%%%%%%%%%%%%%%%%%%%%%%%%%
\begin{figure}[htb!]
\centering
\includegraphics[width=0.5\textwidth]{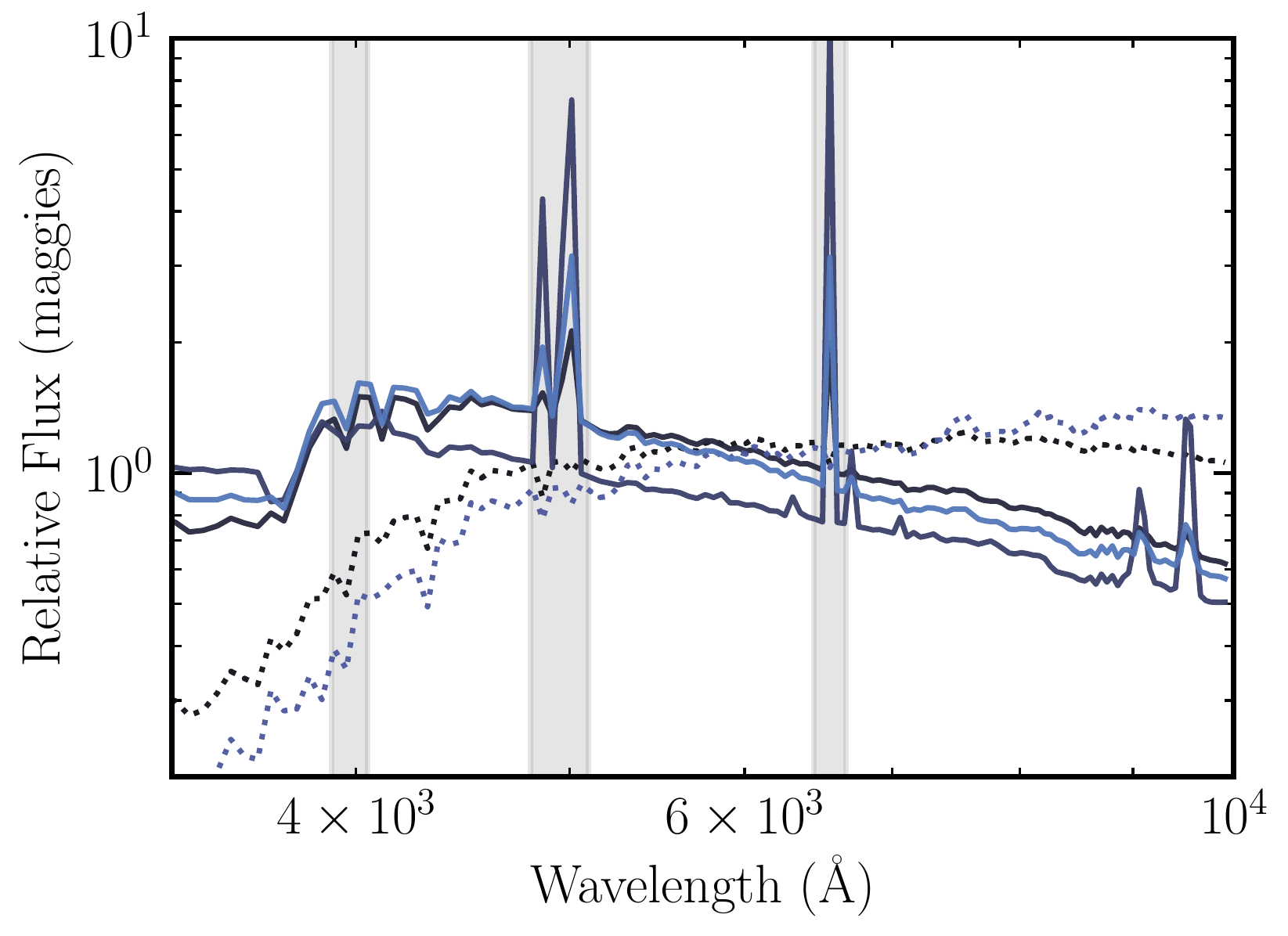}
\vspace{-7mm}
\caption{Five Galaxy SEDs randomly sampled from the training set used in this work, with 5 different values of the parameter vector $\theta$.
The SEDs are normalized to their median: this respects the fact that the shape of the SED is primarily dictated by the stellar metallicity, age and dust properties, while stellar mass is usually correlated with the amplitude of the SED flux.
The SEDs marked by solid lines have nebular emission features (such as the H$\beta$, [OIII]5007\AA\ and H$\alpha$), while dotted lines represent SEDs of galaxies with absorption features and continuum breaks, such as 4000\AA\ .} 
\label{fig:example_seds}
\end{figure}
%%%%%%%%%%%%%%%%%%%%%%%%%%%%%

Fitting a large number of free parameters is computationally expensive. 
A 5-parameter spectral model within a typical SED fitting code -- stellar mass, dust attenuation, metallicity, age -- converges to a best-fit model solution in 2-10 CPU hours. Moreover, each galaxy spectra requires its own separate inference chains. With the advent of spectroscopic surveys that may potentially observe tens of millions of galaxy spectra, the cost of modeling spectra quickly becomes prohibitive. 
Thus, the need of the hour is to quickly and reliably deduce the physical parameters of galaxies in large surveys, as well as to rapidly analyse thousands of spectra from within one galaxy.
%%%%%%%%%%%%%%%%%%%%%%%%%%%%%%%%%%%%

%%%%%%%%%%%%%%%%%%%%%%%%%%%%%%%%%%%%
\begin{table*}
\centering
\noindent\begin{minipage}[b]{\textwidth}
\centering
\caption{Simulated SEDs: Model Description and Prior Range}
\label{table:sed}
\centering
\footnotesize
\begin{tabular}{|l| c|c|}
\hline
\bf{Parameter} & \bf{Description} & \bf{Priors}\\
\hline
$M_{\rm total} (M_{\odot})$ & Total stellar mass formed & Log$_{10}$ Uniform: [10$^{8}$, 10$^{13}$] \\
\hline
$\log(Z/Z_{\odot})$ & Stellar metallicity in units of $\log(Z/Z_{\odot})$ & Uniform: [-2.0, 0.2]\\
\hline
$\tau_{\lambda, 2}$ & Diffuse dust optical depth& Tophat: [0.1, 10.00]\\
\hline
$t_{age}$ & Age of Galaxy (Gyr) & TopHat: [0, 4]\\
\hline
$\tau$ & e-folding time of SFH (Gyr) & Log$_{10}$ Uniform: [0.1, 1.0] \\
\hline
\end{tabular}
\end{minipage}
\end{table*}
%%%%%%%%%%%%%%%%%%%%%%%%%%%%%%%%%%%%

Deep learning applied to galaxy SED fitting allows, in principle, a mapping between an observed SED and the target galaxy's star formation history, with several studies in the last few years alone demonstrating the efficacy of new methodologies \citep{lovell2019,astronn_2019,sedflow2022}. 
These methods allow regression of galaxy parameters, albeit without any uncertainty quantification.

Recent developments in deep learning methods have focused on uncertainty quantification.
For example, Deep Ensembles involve retraining a network many times with different initializations to enable uncertainty quantification of the model outputs. \citep{deepensemble_2021}. Furthermore, with Bayesian Neural Networks (BNNs), deterministic weights of the model are replaced by probability distributions, which allows the model to provide uncertainties of its outputs \citep{bnn_2020} (training a BNN includes assumptions of priors over the network weights and assumes that parameter posteriors can be approximated by well behaved variational distribution e.g. Gaussian distribution or a mixture of multiple Gaussians). 
 
Simulation-based inference  \citep[SBI;][]{cranmer2019} combined with deep learning can mitigate assumptions (e.g., tractable likelihood) that can plague analytic likelihood and posterior modeling, as well as remove computational bottlenecks in statistical calculations. Many astrophysical studies have demonstrated success with SBI in calculating posterior estimation in a rapid manner \citep{Kacprzak_2018, alsing2019, zhang2021, zhao_2021, Huppenkothen2022}.

SBI of galaxy spectra is an exciting opportunity because these models a) do not require the expression of an explicit likelihood, and b) can calculate approximate posterior distributions of galaxy parameters efficiently, allowing for robust uncertainty quantification.
Recent work has shown that photometric SED data can be used with SBI for fast inference, e.g.,  \citealt{sedflow2022}, and \citetext{Robeyns et al. (2022)}.

In this work, we demonstrate a proof-of-concept SBI framework to analyse galaxy spectra and recover posteriors efficiently for a 5-parameter SED model. 
We expect to scale this work to upcoming galaxy spectroscopic surveys -- like the Dark Energy Spectroscopy Instrument \cite{desi2016} and the Roman Space Telescope High-Latitude Survey \cite{romanhls2022} --  and for SED models with more complex and flexible descriptions of galaxy properties.

%%% summary
In Section~\ref{sec: Data}, we describe the simulated data used in this study.
In Section~\ref{sec:methods}, we describe the SBI network architecture and analysis, and in Section~\ref{sec:results}, we share our results and next steps. The fiducial cosmology model used for all distance measurements as well as other cosmological values assumes a standard flat cold dark matter universe with a cosmological constant $(\Lambda$CDM), corresponding to WMAP9 observations \citep{Hinshaw_2013}.

%%%%%%%%%%%%%%%%%%%%%%%%%%%%%%%%%%%%%%%%%%%%%%%%%%%%%%%%%%
\section{Data}
\label{sec: Data}

We use \code{Prospector} \citep{johnson21} to generate simulated SEDs of galaxies.
\code{Prospector} relies on Markov Chain Monte Carlo (MCMC) sampling for stellar population synthesis (SPS) and parameter inference.
It is based on the \code{Python-FSPS} framework, with the MILES stellar spectral library and the MIST set of isochrones
\citep{2010ApJ...712..833C,2017ApJ...837..170L,2013PASP..125..306F,2011A&A...532A..95F,Choi_2016}. 

We generate a training set of 10000 rest-frame SEDs using a 5-parameter model, with a delayed, exponentially declining (i.e., delayed-tau) star formation history. 
The SFH -- star formation rate as a function of time -- is:
\begin{gather}
\operatorname{SFR}(t, \tau) \propto t / \tau * e^{-t / \tau}
\end{gather}

where SFR is the star formation rate, t is the epoch at which the star formation history is being evaluated, and $\tau$ corresponds to the e-folding time in the delayed-$\tau$ SFH model.

%%%%%%%%%%%%%%%%%%%%%%%%%%%%%%%%%%%%
\begin{figure*}[htb!]
%\vspace{-6mm}
\centering
\includegraphics[width=0.95\textwidth]{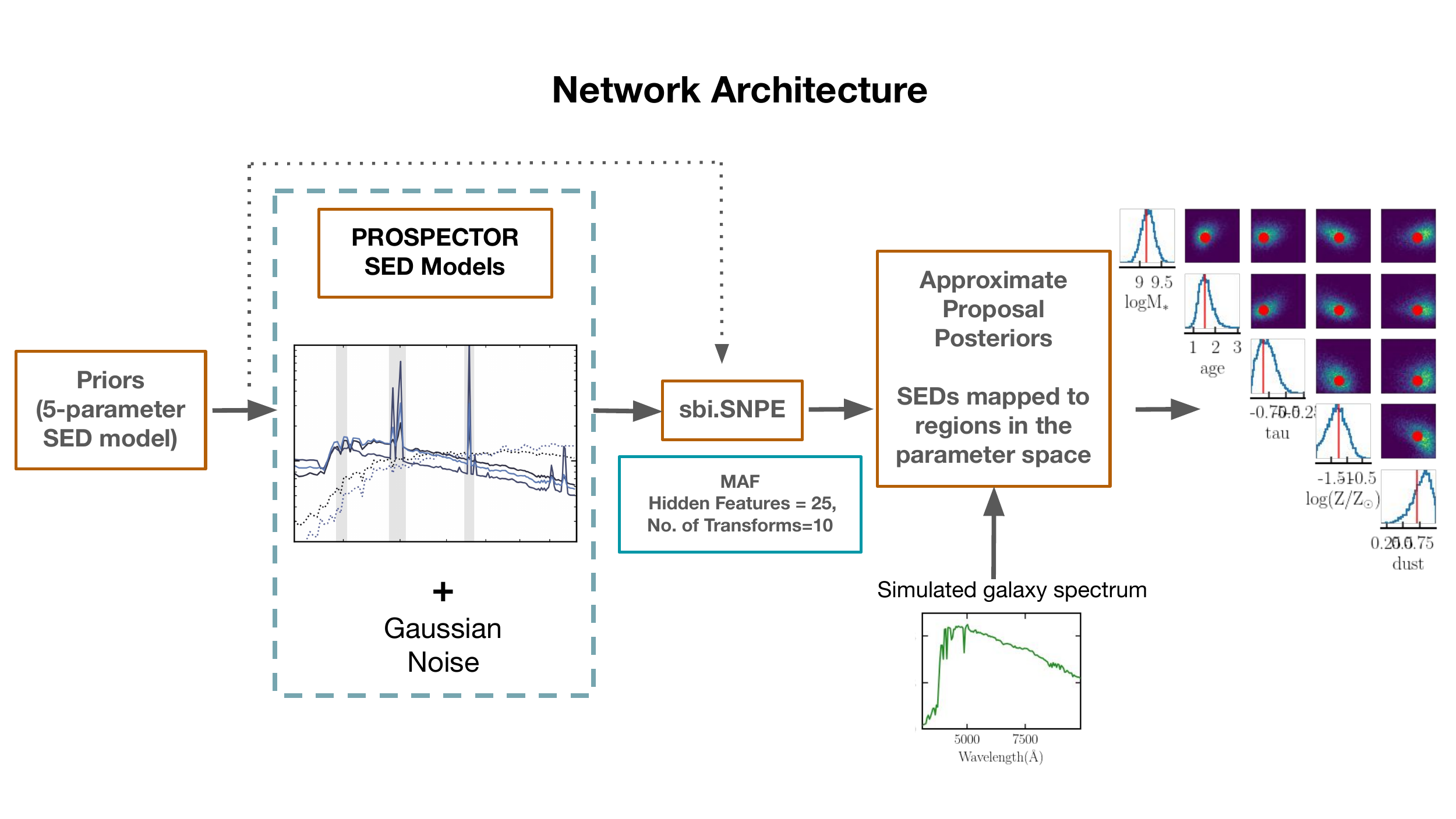}
\vspace{-3mm}
\caption{The architecture used in this work to infer galaxy SED properties with spectroscopic data. We use a 5-parameter model and a training set with realistic spectra, that is trained by an NPE to generate approximate posteriors.}
\label{fig:architecture}
\end{figure*}
%%%%%%%%%%%%%%%%%%%%%%%%%%%%%%%%%%%%

This model incorporates physical priors used in survey studies of galaxy mass assembly (e.g., see \citealt{belli2019}).
We sample the total stellar mass (M$_*$), the stellar metallicity $log(Z/Z_{sol})$, a delayed and exponentially declining SFH with age $t_{age}$, the e-folding time $\tau_{age}$ ($\tau$ from here on) , and dust attenuation ($\tau_{\lambda, 2}$, corresponding to the optical depth of diffuse dust at 5500\AA; from here on referred to as dust.)
Each parameter vector $\boldsymbol{\theta}$ comprises these five parameters.
Our SED model assumes a Kroupa Initial Mass Function (IMF) \citep{Kroupa_2001}. 
Nebular continuum and line emission are also present. 

See Table~\ref{table:sed} for a description of the prior range for each model parameter. See Figure~\ref{fig:example_seds} for 5 example SEDs/spectra from our training set, highlighting the emission and absorption features depending on the type of galaxy (young vs old stellar populations, respectively). 

We smooth and resample the simulated SEDs to resemble a medium-resolution spectroscopic survey using \code{Prospector}'s internal resampling utility.
We use a velocity smoothing parameter ($\sigma_{v}$ (km s$^{-1}$), to account for the contribution of Doppler broadening by stellar velocities, and resolution of the model libraries), and fix it at 350 km$s^{-1}$; this smoothing corresponds to R$\sim$100, similar to a deep galaxy survey conducted with the R$\sim$100 JWST/NIRSpec prism \citep{2017ApJ...836...78Z}. 
This results in a training set with each galaxy SED sampling rest-frame 3750-9500\AA, with 138 flux elements for each SED, which is the data vector $\boldsymbol{x}$ corresponding to each $\boldsymbol{\theta}$.

To map our training set to observations, we add stochasticity to the training set in the form of Gaussian noise, to the level of 5\% of the flux at a given wavelength, representative of real data at signal-to-noise ratio SNR$\sim$20. 
We conduct data augmentation to scale 10000 noise-less spectra in our base training set to 2$\times$10$^6$ spectra with Gaussian noise used in our SBI framework (see Section~\ref{sec:methods}).

We also create an additional test of pectra conduct posterior diagnostics. 
We generate a set of 1000 spectra with noise, from the same parameter prior range.

%%%%%%%%%%%%%%%%%%%%%%%%%%%%%%%%%%%%
\section{Inference Methodology}
\label{sec:methods}

\subsection{Simulation-Based Inference}

Our objective is to calculate posterior distributions $p(\boldsymbol{\theta} |\boldsymbol{x})$ of the galaxy parameters derived from a typical SED analysis, where  $\boldsymbol{\theta}$ is the set of galaxy properties, and $\boldsymbol{x}$ represents the galaxy spectra.  
We do this by training our SBI model on the large stochastically-sampled training set of SEDs described in Section~\ref{sec: Data}. We utilise Neural Posterior Estimation (NPE) \citep{NIPS2016_6aca9700,2019arXiv190507488G}) which relies on neural networks to train on simulated SEDs with realistic noise, and allow us to estimate ``amortized" posterior distributions.

SBI/NPE requires computational time in advance of the actual inference, and evaluates the posterior for different observations without having to re-run inference (this is known as amortization).
This ``amortized'' calculation of posteriors then allows us to infer the posteriors of a ``real'' galaxy with computational time $<$ 1s. 
For more details and examples of amortized neural network-based posterior estimation, see \citet{2019arXiv190507488G} or Section 2 of \citet{sedflow2022}. We provide a short summary below.

NPE uses ``normalizing flows'' \citep{tabak2013} as a density estimator, which employs an invertible bijective transformation to map a complex distribution (i.e., the true posterior distribution in SED model parameter space) to a simpler and faster-to-calculate distribution (often Gaussian, or a combination of Gaussians). 
This results in the calculation of approximate posterior distributions, that are assumed to be a good approximation of the underlying posterior distributions of parameters. 
In particular, we use Masked Autoregressive Flow (MAFs; \citealt{uria2016,papamakarios2017}) incorporated within \code{sbi} (similar to \citealt{sedflow2022}). MAFs perform well in modeling conditional probability distributions, such as posteriors (see Section 2 of \citealt{sedflow2022} for more details). 

%%%%%%%%%%%%%%%%%%%%%%%%%%%%%
\begin{figure}
\centering
\includegraphics[width=0.5\textwidth]{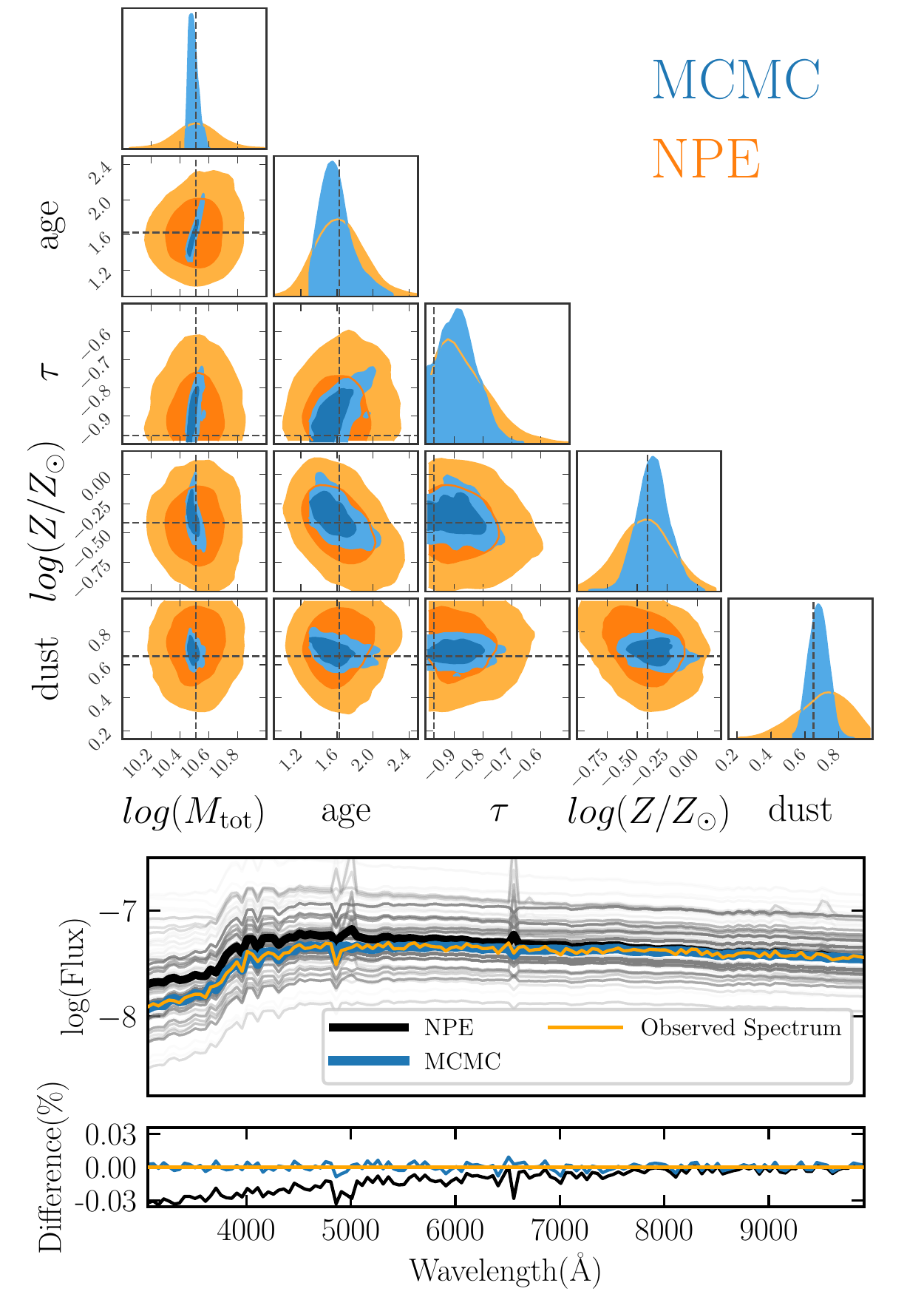}
\caption{(Top) Corner plot for an example galaxy in our test set for a 5-parameter SED model. 
This plot shows pairwise posteriors inferred from the MCMC (blue) and SBI neural posterior estimation (orange) analyses. 
We find that SBI and MCMC recover SED parameters $\boldsymbol{\theta}$ accurately, while SBI constraints for each parameter (for marginalized posteriors) is similar or moderately weaker than MCMC constraints.
(Bottom) Best-fit SEDs from SBI/NPE (black) and MCMC (blue) analyses, and percentage residuals.
Note that the difference in observed spectrum (orange) and NPE best-fit spectrum (black) is $<$ 3\%, smaller than the Gaussian noise applied to each simulated spectrum.}
\label{fig:corner}
\end{figure}
%%%%%%%%%%%%%%%%%%%%%%%%%%%%%

\subsection{This work}

See Figure~\ref{fig:architecture} for a depiction of the SBI architecture used in this work. We use a supervised learning pipeline within an SBI framework via the Macke Lab \texttt{sbi} toolbox  \citep{tejero-cantero2020sbi}. 
To demonstrate a proof-of-concept, we train on 2$\times$10$^6$ simulations (where each simulation is a noise-added version of an SED in our training set) in an NPE framework. We use 25 hidden units and 10 transform layers without an embedding network in this framework. Our model trains on features in the raw simulated data; this model converges after 87 epochs and takes $\sim$ 14 hours to train. 
This analysis generates the set of approximate posterior distributions for our 5 parameter SED model. 
We also test on other combinations of hidden units and transform layers, and choose the above as the fiducial choice with more robust results.

\subsection{Posterior Diagnostics}

We evaluate the results using a variety of statistical and diagnostic mechanisms. 
First, to test the precision of our framework, we compare the recovered SED parameter values with the true values $\boldsymbol{\theta}$ of the parameters from our test set. 
Secondly, to test the health of the calculated posteriors, we also perform posterior predictive checks (PPCs) and simulation-based calibration (SBC) checks \citep{talts2018}.
\cref{PPCs validate that the model SEDs corresponding to the distribution of $\theta$ values in a given posterior fall within the allowed range. 
We do this by cross-checking whether our best-fit model-based spectrum looks similar to observed data $x$ (see Figure \ref{fig:corner}).

SBCs provide a quantitative insight into whether the posterior uncertainties are balanced. 
In this test, we sample $\boldsymbol{\theta}_i$ values from the priors, and simulate observations (using our simulator) from these parameters.
Following this, we perform inference given each of these observations, which generates SBC posteriors of their own.
For a healthy posterior, the SBC ranks of ground truth parameters under the inferred posteriors should follow a uniform distribution (rank plots aid in visually confirming this; see Figure \ref{fig:rank_hist}).}

Finally, we compare our results to an MCMC analysis of representative SEDs from the test set.

\subsection{MCMC Analysis} \label{sec:mcmc}

We fit the same 5-parameter SED model to representative galaxy SEDs in the test set using the inference framework \code{Prospector}, which calculate Markov-Chain Monte Carlo (MCMC)-based posterior distributions. 

\cref{We assume the following likelihood function:
$$  \ln\mathcal{L}_{\textrm{spec}}(\boldsymbol{x, \theta, \sigma}) =  \sum_{\mathbf{i}=1}^{\mathbf{n}} \ln \left(\frac{1}{\sqrt{2 \pi} \sigma_{\mathbf{i}}}\right)-\frac{1}{2} \sum_{\mathbf{i}=1}^{\mathbf{n}} \frac{\left(\mathbf{x}_{\mathbf{i}}-\mathbf{m(\theta)}\right)^{2}}{\sigma_{\mathbf{i}}^{2}}\ $$ 

\noindent where ($x, \sigma_{\mathbf{i}}$) are n independent spectral flux elements assumed to be drawn from a Gaussian distribution, and $\mathbf{m(\theta)}$ corresponds to the model spectrum for a given parameter set $\boldsymbol{\theta}$. }

%%%%%%%%%%%%%%%%%%%%%%%%%%%%%
\begin{figure*}[htb!]
\centering
\includegraphics[width=1.0\textwidth]{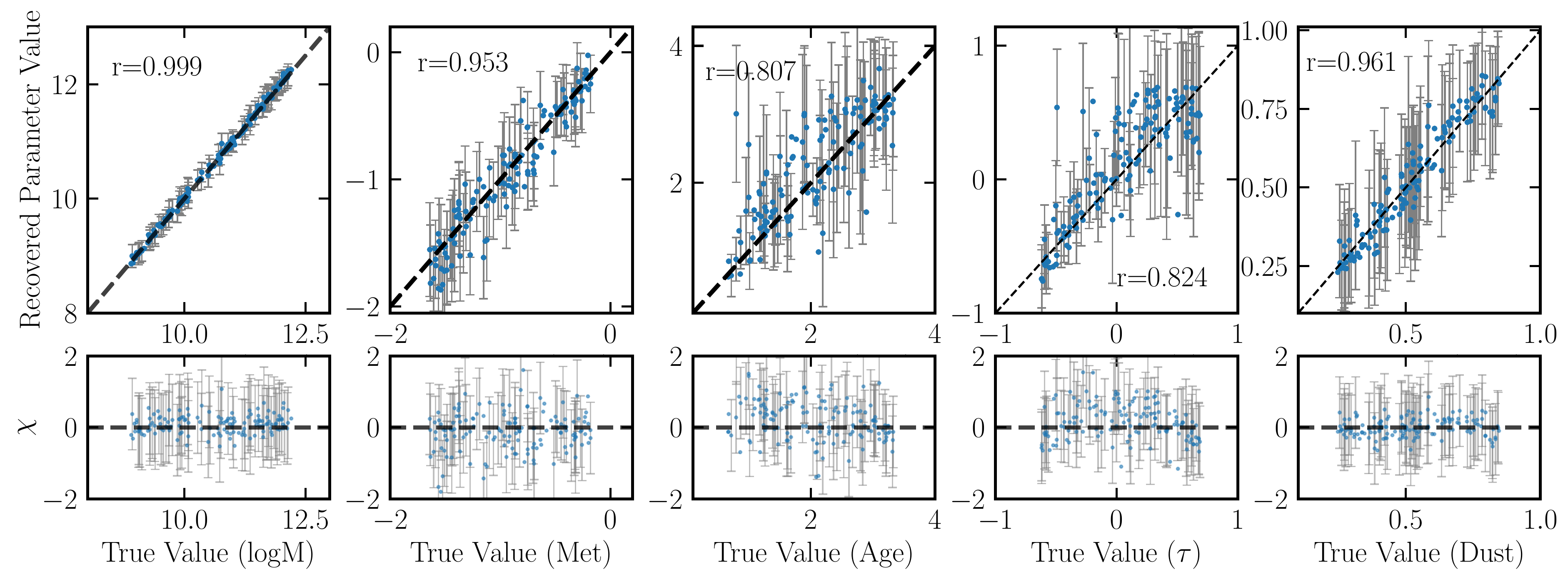}
\caption{(Top) True vs. recovered values for SED parameters in the test set of 1000 spectra, sampled from the 16-84th percentile range of priors in this study. This demonstrates the accuracy of the predicted models across the entire range of priors. Note that only every alternate errorbar is plotted for visual clarity in the plot. (Bottom) Goodness-of-fit ($\chi$) plots for each parameter, with errorbars corresponding to a value of 1. } 
\label{fig:parameters_true_recovered}
\end{figure*}
%%%%%%%%%%%%%%%%%%%%%%%%%%%%%

We use the same $\boldsymbol{\theta}$ prior range and shape as the SBI/NPE analysis, in order to calculate the posterior $p(\boldsymbol{\theta} |\boldsymbol{x})$.
We use \code{emcee} \cite{2013PASP..125..306F} to conduct the MCMC posterior sampling with 128 walkers, 128 iterations and a burn-in with the step set [4096,4096,2048,512].

\cref{Note that non-Gaussian or correlated uncertainties are seen in spectral datasets (e.g., magnitude upper limits in the case of non-detections), which are not accurately captured by the above likelihood, making a "likelihood-free inference" like SBI the ideal choice for this analysis. The results from the SBI analysis, posterior diagnostics, and MCMC comparison are shown in Section~\ref{sec:results}.}

\subsection{Computing Resources}

For our SBI analysis, we use the Python 3 Google Compute Engine backend (with the CPU processor AMD EPYC 7B12), which for our network architecture takes $\sim$14 CPU hours to train 2$\times$10$^6$ simulated spectra with noise. For every subsequent posterior estimation, this setup takes $\sim$0.3s. 

For our serialized MCMC calculations, we utilise \code{Prospector} runtime on a 2.7 GHz Quad-Core Intel Core i7 processor, which takes $\sim$14 CPU hours to converge. 

%%%%%%%%%%%%%%%%%%%%%%%%%%%%%%%%%%%%%%%%%%%%%%%%%%%%%%%%%%

\section{Results}
\label{sec:results}

In this proof-of-concept analysis, we test out to set whether an SBI framework can train on realistic noisy galaxy spectra to estimate amortized posteriors robustly. 
To test the efficacy of our SBI framework, we use a test set of 1000 randomly sampled SEDs from our prior range (see Table~\ref{table:sed} and Section~\ref{sec: Data}). 

One such result is shown in the top panel of Figure~\ref{fig:corner}, for a galaxy with logM$_{tot}$= 10.51 (M$_{\odot}$),  $log(Z/Z_{\odot})$ = -0.41, age = 1.63 Gyr, $\tau$ = 0.11 Gyr, and dust = 0.65 (a metal-poor dusty galaxy). 
We plot pairwise posterior distributions estimated from both the MCMC (in blue) and SBI (or neural posterior estimation; NPE, in orange) in order to compare constraints across the 5-parameter SED model.
The truth values are plotted with black dotted lines.
In the bottom panel of Figure~\ref{fig:corner}, we show the maximum a posteriori (MAP) SED models and model residuals from the SBI/NPE (black) and MCMC (blue) analyses.
Also overplotted are 1000 randomly sampled SEDs from the posteriors (in grey).

We find excellent agreement between the median values of parameters across MCMC and SBI/NPE posteriors (when marginalized over other parameters); these values are also accurate relative to the true parameter values $\boldsymbol{\theta}$. 
We also observe that the age and metallicity constraints are similar in both analyses for this test galaxy, while MCMC mass and dust estimation is more precise relative to SBI/NPE. 
This is the first demonstration that the proof-of-concept analysis presented here is effective at recovering galaxy SED parameters with spectroscopic observations.

%%%%%%%%%%%%%%%%%%%%%%%%%%%%%
\begin{figure*}[htb!]
%\vspace{-6mm}
\centering
\includegraphics[width=1.0\textwidth]{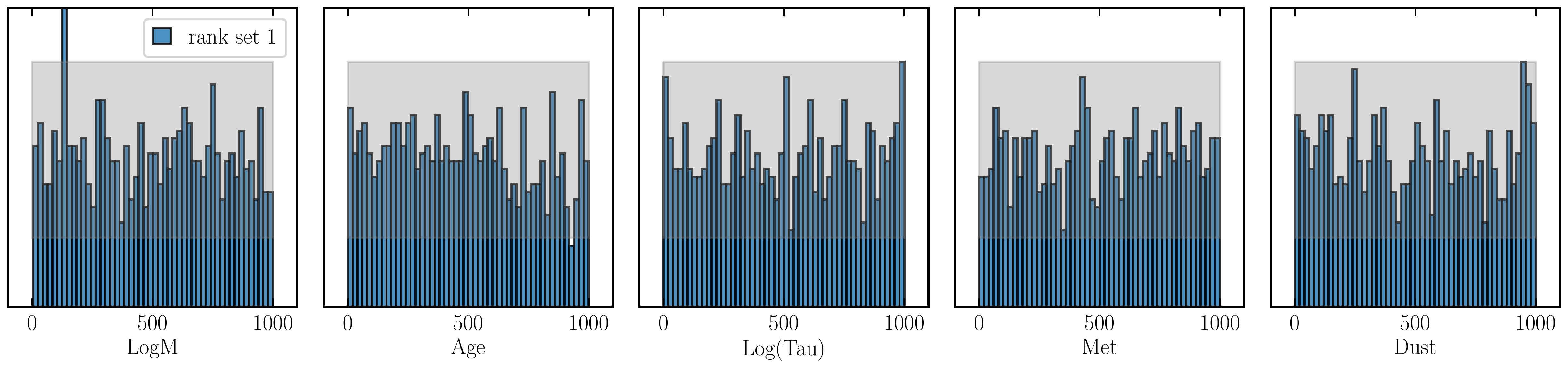}
\caption{Simulation-Based Calibration rank plots for the SBI/NPE analysis. Each subplot corresponds to a parameter in the SED model. The grey region corresponds to the 95\% confidence interval of a uniform distribution, which our parameter rank distributions follow.}
\label{fig:rank_hist}
\end{figure*}
%%%%%%%%%%%%%%%%%%%%%%%%%%%%%
\begin{figure*}[htb!]
%\vspace{-6mm}
\centering
\includegraphics[width=0.8\textwidth]{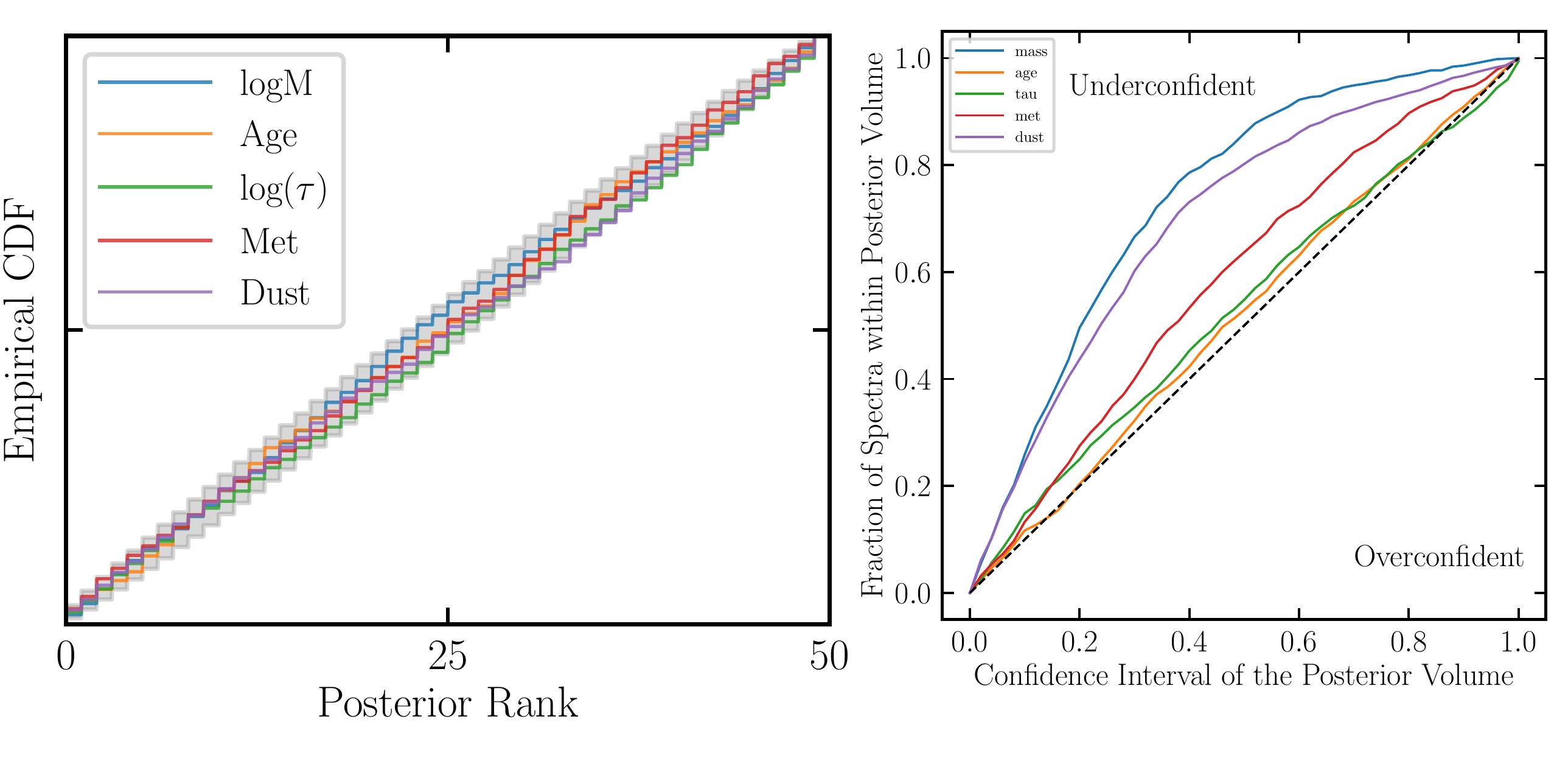}
\vspace{-5mm}
\caption{(Left) The cumulative density function (CDF) of posterior ranks for each parameter in our SBI/NPE analysis relative to the 95\% confidence interval of a uniform distribution (grey). 
(Right) Probability coverage plot for each model SED parameter. 
A well-calibrated posterior estimator will produce curves that closely follow the dashed line (See Section~\ref{sec:results} for additional information).}

\label{fig:rank_cdf}
\end{figure*}
%%%%%%%%%%%%%%%%%%%%%%%%%%%%%

On running inference on a sample of 1000 test galaxies sampled from the 16th-84th percentile range of priors in this study, we find accurate recovery of SED parameters.  
See Fig.~\ref{fig:parameters_true_recovered} for a comparison between true and recovered values of each parameter, as well as $\chi$ plots to show goodness-of-fit across the simulated spectroscopic dataset. 
The recovered values here are the 50th percentile parameter values, and the uncertainties correspond to confidence intervals between 16th-84th percentile in the posteriors.
Here, we demonstrate accurate and precise parameter recovery across the majority of the prior range, specifically for stellar mass (logM), while the constraints on metallicity and age are seen to be wider and less precise. 
We also note that in our analysis, the recovered parameters are the most biased at the edges of the prior ranges, which indicates that the underlying posterior distribution is not being captured in these parameter ranges.
For example, the 16th, 50th and 84th percentile of parameter values for a given galaxy are not accurate descriptors of the underlying posteriors near the edges of the prior range. 
This can be potentially solved by training on a spectroscopic dataset sampled from a prior range marginally wider than the target spectroscopic survey. 
We also find no substantial difference in the quality of parameter recovery for star forming galaxies (emission line galaxies; median values of the age parameter $<0.75\        \mathrm{Gyr}$), or quiescent systems (galaxies with spectra containing strong absorption line indices; median values of the age parameter $ > 2.5\ \mathrm{Gyr}$).

We also run extensive PPC and SBC checks to test the accuracy and precision of our SED parameter values, where we find that the posterior distributions in this analysis are well converged. See Figure~\ref{fig:rank_hist} for rank distributions for each parameter — a healthy posterior follows uniform distribution (non-uniform distributions indicates a poorly calibrated posterior; \citealt{talts2018}). 
This demonstration of well-calibrated uncertainties in our SBI analysis is confirmed in Figure~\ref{fig:rank_cdf}, where we plot the rank cumulative distribution function (CDF) on the left panel. 
In both figures, the grey region corresponds to the 95\% confidence interval of a uniform distribution, which our parameter rank distributions follow.

The right panel of Figure~\ref{fig:rank_cdf} shows the probability coverage curve of each SED parameter $\boldsymbol{\theta}$. 
The principle behind a probability coverage plot is as follows: 
a well-calibrated posterior estimator will produce - for an ensemble of SEDs in the test set - parameter uncertainties that accurately reflect the true underlying uncertainty in the ensemble, e.g. a 68\% posterior volume will contain 68\% of the true SED parameter values of the test ensemble.
Generalizing from this, by plotting the fraction of SED parameters in the test set which fall within the  posterior volume as a function of the posterior volume, we obtain curves such as those seen in the right panel of Figure~\ref{fig:rank_cdf}. 
A well-calibrated estimator will produce probability coverage curves that closely trace the diagonal dashed line, while posterior estimates that are under-confident (i.e. over-estimate uncertainties) will produce curves that fall on the upper-left part of the plot.  

For our 5-parameter model, we see that the NPE has fairly well-calibrated uncertainty predictions for the age and $\tau$ parameters. 
On the other hand, it tends to over-predict the posterior uncertainties for three parameters — stellar mass, dust and metallicity. 
This is consistent with the results in Fig.~\ref{fig:parameters_true_recovered} — in the goodness-of-fits plots, the scatter in the differences between the predicted and true parameters values across the test set is smaller than what their error bars would suggest. 
However, we are encouraged by the fact that the NPE analysis a) accurately predicts the median best-fit values across the test set for those 3 parameters, and b) in most scientific applications, over-predicting the uncertainties in an analysis is preferable than the alternative. 
We aim to continue to further improve the posterior uncertainty calibrations in future work. 

%\subsubsection{Speed Comparisons with MCMC}

We also demonstrate here a significant improvement in inference speeds compared with MCMC inference. 
As mentioned above, the SBI/NPE model uses 25 hidden units and 10 transform layers: this computation takes $\sim$14 hours to train on a CPU, and $\sim$0.3s per posterior estimation thereafter (MCMC calculation for a single galaxy takes $\sim$24 hours in our setup). 
Accounting for the cost of training, we can effectively infer accurate posteriors of 0.9-1.2$\times$10$^5$ galaxy spectra via SBI/SNPE at the cost of 1 MCMC-based fit. 
This demonstrates that the amortization of posterior estimation in SBI with accurate recovery of SED parameters is the biggest advantage of our proof-of-concept analysis.

\subsection{Caveats}

\cref{This work presents SNPE analyses on spectra generated using a) empirical templates within FSPS (MILES spectral libraries, and MIST isochrones; \citealt{2011A&A...532A..95F,Choi_2016}), and b) with an assumption of uniform (or nearly uniform) SNR and Gaussian noise across wavelength and for each galaxy spectrum.
These assumptions have impact on both SNPE and MCMC inference of galaxy parameters.}

\cref{For example, the age-metallicity-dust degeneracy is a systematic that is limited by the information content of the templates, and the wavelength range sampled by the spectra that may or may not contain discriminating information. 
This affects both MCMC and SNPE analyses, and is expected to be mitigated with upcoming spectroscopic surveys of statistical samples of galaxies (DESI, or the Prime Focus Spectrograph; \citealt{pfs2022}) as well as simulation studies such as UniverseMachine \cite{2013ApJ...770...57B} and FIRE \cite{2018MNRAS.478.1694M}.}

\cref{In addition, SED inference using spectra is only weakly impacted by small ($<10-20\%$) variations in SNRs across wavelengths (especially in the stellar continuum portions of a given spectrum.
Several studies analysing quiescent galaxy spectra  \cite{carnall2019, khullar_2022a, tacchella_2022_fsel} as well as photometric SED fitting and parameter recovery \cite{10.1086/679742, Leja_2017} seem to point towards this trend.
Star-forming galaxies (with strong emission lines) have wavelength regions with peaked SNRs, that improves constraints on parameters such as instantaneous star formation rates (SFRs) and ages.
This biases our inferences in favor of young stellar populations with emission line spectra, albeit weakly. 
In Figure \ref{fig:parameters_true_recovered}, we observe that both older and younger stellar populations are recovered with similar precision.
Finally, we expect spectroscopic surveys like DESI, PFS and the Roman High-Latitude Survey to improve this systematic uncertainty, as they attempt to reach nearly uniform SNRs across wavelength.}

%%%%%%%%%%%%%%%%%%%%%%%%%%%%%

\section{Conclusion and Next Steps}
% \vspace{1mm}

In this work, we demonstrate a proof-of-concept for amortized neural posterior estimation with an SBI framework, that utilizes simulated low/medium-resolution galaxy spectra. 
This is the first-of-its-kind demonstration of this technique on spectra, that will enable precise and rapid estimation of galaxy parameter posteriors for billion-galaxy surveys. 
We also show here a significant improvement in inference speeds, while maintaining accuracy in the recovery of parameters, with precision comparable or moderately weaker than MCMC constraints.

While this work focuses on using an SBI framework to train on galaxy spectra directly (without any summary statistics or embedded nets), we wish to scale this analysis with GPU-processing on suitable summary statistics (Khullar et al. in prep, 2022).
Moreover, the combination of highly complex SED models \citep{Leja_2019,suess_psb_2022} and high-resolution spectroscopy will enable precise constraints on star formation histories of galaxies.
This is especially true in the era of JWST \citep{2022arXiv220713860N,2022arXiv220711135L,2022arXiv220712446L,2022arXiv220710655S} and Roman Space Telescope (High Latitude Survey; \citealt{roman_hls_2022}), where SED analysis of systematic spectroscopic surveys will be bolstered with an SBI framework.

\cref{Finally, when using simulation-trained SBI models on future survey data, it is important to consider possible performance issues that will arise from small differences between simulated and real data (due to approximations, unknown physics or computational constraints, imperfect simulation of noise and other observational effects). 
The drop in performance of simulation-trained models that are applied to real data is a known issue that affects all deep learning models.
Mitigation of these problems is an active area of research, which already led to the development of a broad group of methods called \textit{Domain Adaptation}~\citep{C2017,WD2018}. 
These methods allow deep learning model to learn the features shared between simulated and real data and use only these features for inference. 
This leads to better alignment of the two data distributions in the latent space of the deep learning model, which leads to improved performance~\citep{CK2021,CK2022}. 
In future work, we will include domain adaptation methods in our SBI frameworks. }

%%%%%%%%%%%%%%%%%%%%%%%%%%%%%%%%%%%%
%%%%%%%%%%%%%%%%%%%%%%%%%%%%%%%%%%%%
\section*{Acknowledgments}
%%%%%%%%%%%%%%%%%%%%%%%%%%%%%%%%%%%%
%%%%%%%%%%%%%%%%%%%%%%%%%%%%%%%%%%%%
The authors thank Alexander Ji, Egor Danilov, Michael D. Gladders for their comments and feedback in the planning and analysis of this work. 
GK thanks the URA Visiting Scholars Program, 2021, for funding this work through graduate student salary support. 

\cref{The authors are grateful to the reviewers of the journal MLST for their extremely thoughtful and helpful comments; their efforts and feedback have substantially improved the quality of the manuscript.}

This manuscript has been supported by Fermi Research Alliance, LLC under Contract No.\ DE-AC02-07CH11359 with the U.S.\ Department of Energy (DOE), Office of Science, Office of High Energy Physics. 

The authors of this paper have committed themselves to performing this work in an equitable, inclusive, and just environment, and we hold ourselves accountable, believing that the best science is contingent on a good research environment.

We acknowledge the Deep Skies Lab as a community of multi-domain experts and collaborators who have facilitated an environment of open discussion, idea-generation, and collaboration. This community was important for the development of this project.

\subsection*{\it{Author Contributions}}

G.~Khullar: \textit{Conceptualization, Data curation, Formal analysis, Investigation, Methodology, Project administration, Resources, Software, Visualization, Writing of original draft}; 
B.~Nord: \textit{Conceptualization, Investigation, Methodology, Project administration, Resources, Acquisition of financial support for this publication, Supervision, Writing (review \& editing)};
A.~\'Ciprijanovi\'c: \textit{Investigation, Methodology, Analysis, Project administration, Resources, Software, Supervision, Writing (review \& editing)};
J.~Poh: \textit{Methodology, Analysis, Resources, Writing (review \& editing)};
F.~Xu: \textit{Methodology, Resources}.

%%%%%%%%%%%%%%%%%%%%%%%%%%%%%%%%%%%%%%%%%%%%%%%%%%
%%%%%%%%%%%%%%%%%%%%%%%%%%%%%%%%%%%%%%%%%%%%%%%%%%
\section*{Data and Code Availability Statement}
%%%%%%%%%%%%%%%%%%%%%%%%%%%%%%%%%%%%%%%%%%%%%%%%%%
%%%%%%%%%%%%%%%%%%%%%%%%%%%%%%%%%%%%%%%%%%%%%%%%%%

%The data that support the findings of this study are openly available at the following URL/DOI:
% \href{https://doi.org/10.5281/zenodo.5514180}{https://doi.org/10.5281/zenodo.5514180}. 

The code and dataset used to perform the experiments presented in this paper is openly available in our GitHub repository: 

\noindent\href{https://github.com/gkhullar/digs_sbi.git}{https://github.com/deepskies/digs\_sbi}

Access to the repository and data are available upon publication.

%%%%%%%%%%%%%%%%%%%%%%%%%
%%%%% Bibliography %%%%%%
%%%%%%%%%%%%%%%%%%%%%%%%%

\bibliographystyle{model2-names.bst}
\bibliography{main}

\end{document}